\documentclass{WileyMSP-template}

\begin{document}

\pagestyle{fancy}
%\rhead{\includegraphics[width=2.5cm]{VCH-logo.bmp}}

\title{Comparative analysis for the behavior of beryllium and magnesium crystals at ultrahigh pressures}

\maketitle

% Author: Please give full first and last names for authors and include * after the name of all corresponding authors

\author{N. A. Smirnov}

% Dedication

\dedication{}

% Affiliations: Please provide adacemic titles (Prof. or Dr.) for all authors where applicable, and include an institutional email address for all corresponding authors
\begin{affiliations}
Dr. Nikolai A. Smirnov\\
Federal State Unitary Enterprise, Russian Federal Nuclear Center - Zababakhin All-Russian Research Institute of Technical Physics, 456770, Snezhinsk, Russia\\
Email Address: deldavis@mail.ru
\end{affiliations}

% Keywords: Please provide a minimum of three and a maximum of seven keywords, separated by commas


% Abstract should be written in the present tense and impersonal style (i.e., avoid we), and be at most 200 words long
\begin{abstract}

The paper presents \emph{ab initio} calculation results on the structural phase stability of beryllium and magnesium crystals under high and ultrahigh pressures (multi-terapascal regime). Magnesium is shown to undergo a number of structural transformations which markedly reduce the crystal packing factor. As for beryllium, its high-pressure body-centered cubic phase remains stable even under ultrahigh pressures. Changes in the electronic structure of Be and Mg crystals under compression are analyzed and some interesting effects are revealed. Specifically, a narrow band gap appears in the electronic structure of magnesium under pressures above 2.5 TPa. For the metals of interest, $PT$-diagrams are constructed and compared with available experimental and theoretical results from other investigations.

\end{abstract}

% Text: Please use section headings and subheadings as specified below. For communications, all section headings apart from Experimental Section should be removed
% Please make the first reference to a display item bold: \textbf{Figure 1}
% Do not abbreviate Figure, Equation, etc.; display items are always singular, i.e., Figure 1 and 2.
% Equations are always singular, i.e., Equation 1 and 2, and should be inserted using the {equation} environment, not as graphics
% Please do not use footnotes in the text, additional information can be added to the Reference list.

\section{Introduction}

Advances in experimental methods for the shock loading of materials allow researchers to get information on their structure and dynamic properties under pressures to 1 TPa and above in laboratory conditions \cite{A01,A02,A03,A04}. Of special interest here are quasi-isentropic (ramp) compression experiments which allow determination of the crystal structure of compressed material \cite{A01,A02,A03}. With this information it becomes possible, for example, to judge the state of material inside massive planets within our Solar system and beyond. Knowledge of the processes that occur under such extreme conditions allows researchers to correctly simulate and predict the state of matter at high pressures and temperatures.

In this paper we consider structural changes in the light alkaline-earth metals beryllium and magnesium under high and ultrahigh pressures. Beryllium, the second lightest metal in the periodic table, exhibits a number of unique properties. Its electronic density of states (DOS) greatly differs from the nearly-free-electron DOS and has a pseudogap near the Fermi level \cite{A05}. Under ambient conditions Be has the hcp structure, a very high Debye temperature, a small Poisson's ratio, and a lattice ratio, $c/a$, much smaller than the ideal value. Beryllium is particularly intriguing due to its hardly observable polymorphism under high pressures and temperatures. So, at the atmospheric pressure and a temperature above 1.5 kK, its structural transition to the bcc phase before melting was observed in early X-ray experiments \cite{A06,A07}. It was later shown \cite{A08} that under the increasing pressure the hcp-bcc boundary on the $PT$-diagram had a negative slope. New theoretical studies \cite{A09,A10} based on calculations from first principles, suggest that at high temperatures ($T$$>$1 kK) and moderate pressures ($\leq$10 GPa), the phase diagram of beryllium exhibits the so-called bcc pocket that originates at $P$$=$0 and lies between the hcp stability region and the melting curve. But recent experiments on Be compression on the diamond anvil cell \cite{A11} did not find any signs of this cubic structure on the phase diagram. Moreover, results of \emph{ab initio} calculations \cite{A09,A12,A13,A14,A15} clearly show that the bcc structure should also be most energetically preferable at high pressures, above 100 GPa, where another hcp-bcc boundary exists. This boundary has a negative slope giving at $P$$\sim$180 GPa and $T$$\sim$4.5 kK the triple hcp-bcc-liquid point \cite{A12}. However the efforts taken to prove the presence of the hcp-bcc boundary at high pressures and temperatures have been a success neither in static \cite{A11} nor in dynamic \cite{A16} experiments. Calculations from first principles \cite{A14,A17} show that no other structural transformation occurs in beryllium up to 1 TPa.

Unlike beryllium, magnesium does not exhibit such peculiar physical properties at low pressures. Under ambient conditions it also has the hcp structure. Its electronic density of states is similar to the nearly-free-electron DOS for simple metals \cite{A05,A18}. Its compression results in an hcp-bcc structural transformation at $P$$\approx$50 GPa and $T$$=$300 K well detectable in experiment \cite{A19,A20} and convincingly reproducible in \emph{ab initio} calculations \cite{A21,A22,A23}. It was a surprise to detect in experiment its transition to the dhcp structure under moderate pressures about 10 GPa and temperatures above 1 kK \cite{A24}. But the authors of later experiments \cite{A20} could not identify the observed structure surely. The region of this phase on the $PT$-diagram seems to be limited by a narrow area (5-10 GPa) along the melting curve; above 12 GPa it is no longer detectable. Mg compression to 211 GPa at room temperature did not reveal any other structural transformations in static experiments \cite{A20}.

\emph{Ab initio} calculations predict a more interesting, compared to beryllium, structural behavior of magnesium at pressures above 200 GPa. As suggested in early papers \cite{A25,A26}, at least one more structural transformation is expected to occur at a pressure of several hundred GPa. More precise evidence of structural transformations in Mg was reported in recent works \cite{A22,A27} where some transformations at pressures to 1 TPa and slightly above were predicted with random structure search algorithms. So, a transition to the fcc structure is expected to occur in Mg under pressures to about 0.46 TPa. The further increase of pressure indicates that the simple hexagonal (sh) packing of atoms becomes energetically most favourable (at $P$$\approx$0.75 TPa), and the simple cubic (sc) one does at pressures about 1 TPa \cite{A27}. The authors of paper \cite{A22} calculated hcp, bcc, fcc and sh phase boundaries at elevated temperatures but they limited themselves by a maximum of 2 kK and did not try to calculate the melting curve. It is also said in Refs. \cite{A22,A27} that the high-pressure fcc, sh, and sc phases are the so-called electride structures with electron 'blobs' in the interstitial region, i.e., they have non-nuclear maxima in the electronic density. All of this makes magnesium resemble another alkaline-earth metal, calcium, which also has a high-pressure electride simple cubic phase \cite{A28,A29,A30}. From all the above we can see that under high pressures magnesium becomes anything but a trivial metal and demonstrates some exotic features. Note that \emph{ab initio} calculations \cite{A31} show that non-nuclear maxima are also present in the electronic density of beryllium at the normal specific volume.

In the present work we study the structural behavior of beryllium and magnesium crystals under pressure. Unlike previous works \cite{A22,A23,A27} where the pseudopotential method was employed, here the full-potential method (FP-LMTO) is used for structural stability analysis. Changes in the band structure of beryllium and magnesium under compression are compared and their effects are discussed. Some parallels are drawn with other alkaline-earth metals and not only with them. The calculated phase boundaries of beryllium and magnesium are presented in $PT$-coordinates. The position of melting curves is determined and calculated results are compared with other theoretical works and experiments.

\section{Details of Calculations}

In this work, calculations were done with the full-potential all-electron linear muffin-tin orbital method FP-LMTO implemented in the LmtART code \cite{A32}. Phonon spectra calculations for the metals of interest are based on linear response theory. The valence electrons for beryllium are its all 4 electrons, and 2$s$, 2$p$, and 3$s$ for magnesium. Our test calculations for Mg show that the taken number of valence electrons is quite enough to ensure a good accuracy of calculation to maximal compressions considered ($V$/$V_0$$=$0.07). In our case, the 1s Mg state can be ascribed to core states. The crystal structures under study include not only experimentally observed hcp and bcc, but also face-centered cubic (fcc), double hexagonal closed-packed (dhcp), simple hexagonal, simple cubic, and tetragonal $\beta$-tin ($\beta$-Sn) ones. The last phase was included because it is experimentally observed in compressed calcium at low temperatures \cite{A33} and can potentially compete with the other phases of Be and Mg.

\begin{figure}
  \includegraphics[width=\linewidth]{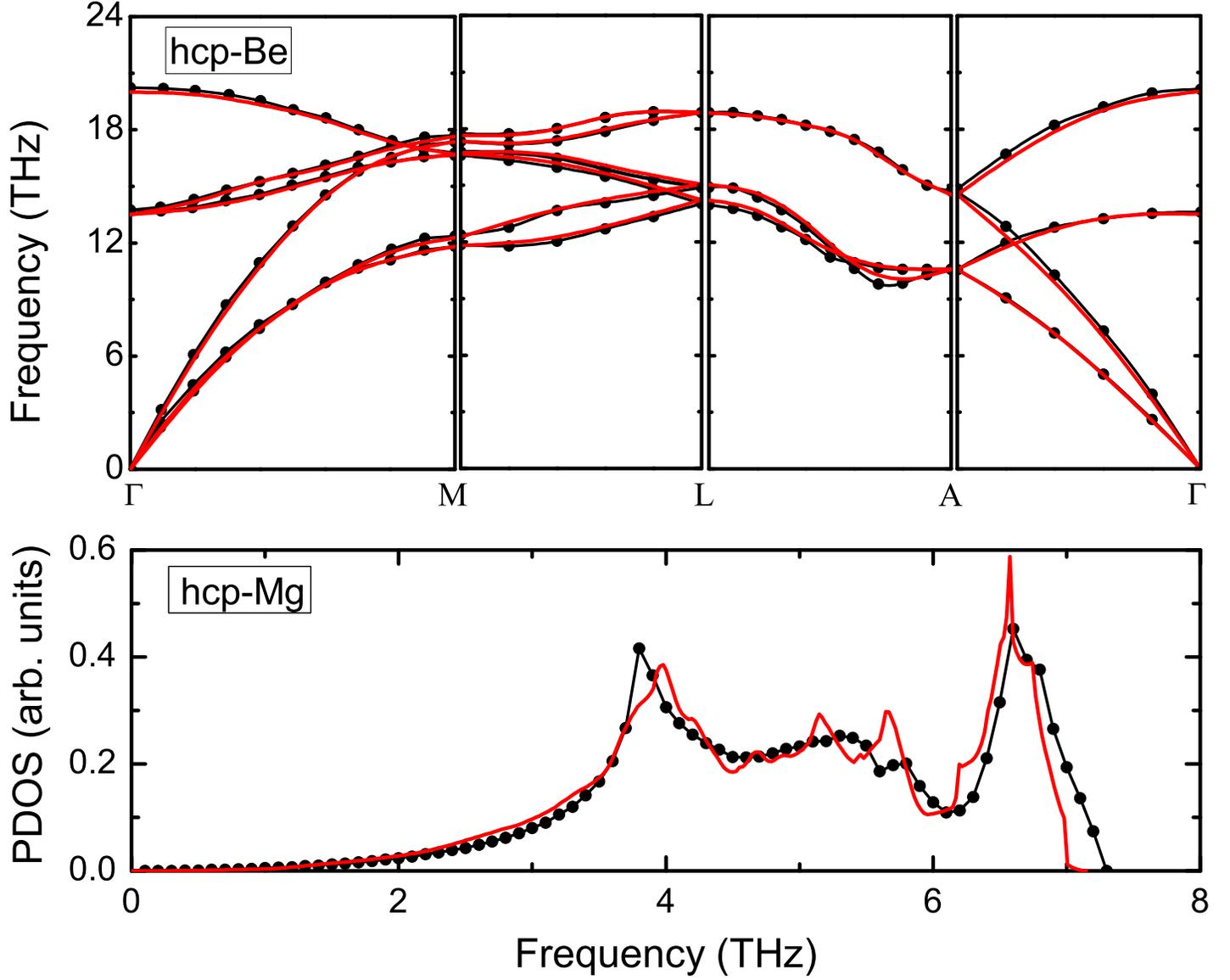}
  \caption{Phonon spectra of hcp Be (upper panel) and hcp Mg (lower panel) from calculation ($V$/$V_0$$=$1) and experiment \cite{A40}. Phonon frequencies in different directions of the Brillouin zone are shown for Be, and the phonon density of states for magnesium. The red line shows calculation at $T$$=$0 K, and the line with black dots shows experimental results at room temperature.}
  \label{fig1}
\end{figure}

In order to achieve high accuracy of first-principles calculations, a careful selection of the internal parameters of the method was carried out. An improved tetrahedron method \cite{A34} was used for integration over the Brillouin zone. The $\vec{k}$-meshes were as follows: 30$\times$30$\times$30 for all cubic structures, 30$\times$30$\times$24 for hcp, 30$\times$30$\times$20 for dhcp, and 24$\times$24$\times$30 for sh and $\beta$-Sn. The cutoff energy for representing the basis functions as a set of plane waves in the interstitial region depended on the magnitude of compression. At the equilibrium specific volume $V$$=$$V_0$ it was 850 eV. The basis set included MT-orbitals with moments to $l^b_{max}$$=$2. Potential and charge density expansions in terms of spherical harmonics were done to $l^w_{max}$$=$6. The $c/a$ ratios of tetragonal and hexagonal structures were always optimized. The internal FP-LMTO parameters such as the linearization energy, tail energies, etc. were chosen using an approach similar to that one used in Ref. \cite{A35}. Calculation parameters including the exchange-correlation functional were selected so as to reproduce best the ground state properties and phonon spectra of the two metals under study. The PBE functional \cite{A36} was taken for both metals. The equilibrium specific volume of Be and Mg under ambient conditions was reproduced no worse than 1\% compared to experiment. Pressure versus volume was determined by differentiating an analytical expression that approximates the calculated dependence of $E$ on $V$. The dependence $E(V)$ was approximated with a formula by Parsafar and Mason \cite{A37}. Crystal internal energy versus relative specific volume $V/V_0$ was calculated for values within intervals from 1.05 to 0.04 for Be and to 0.07 for Mg. Phonon spectra were calculated in the same intervals. Phonon frequencies were determined using meshes of $\vec{q}$-points which measured 10$\times$10$\times$10 for cubic structures, 10$\times$10$\times$6 for close-packed hexagonal phases, and 8$\times$8$\times$10 for sh. The contribution of lattice vibrations to free energy was determined in a quasiharmonic approximation (QHA) \cite{A38} with use of the calculated phonon spectra. The well-known Lindemann criterion was used to evaluate the melting curve. The procedure of its calculation is described in Ref. \cite{A39}. It should be noted that the Lindemann criterion is a rather rough estimate for determining melting curves of materials. The main drawback of this criterion is that it treats the melting process as a lattice instability which occurs at a certain value of the displacement of atoms from equilibrium positions. But this approximation is not always to work well. Although for some materials with simple structures \cite{A35,A39} the Lindemann criterion helps estimate the melting curve quite accurately, for more complex crystals it fails to adequately reproduce it \cite{A39a}. In our calculations, for beryllium the Lindemann constant is equal to 0.109, and for magnesium it is 0.117.

The accuracy of calculated phonon frequencies is demonstrated in Figure~\ref{fig1} that shows the corresponding spectra of beryllium and magnesium calculated in this work in comparison with experimental data. They are seen to agree well.

\section{Results and Discussions}

\begin{figure}
  \includegraphics[width=\linewidth]{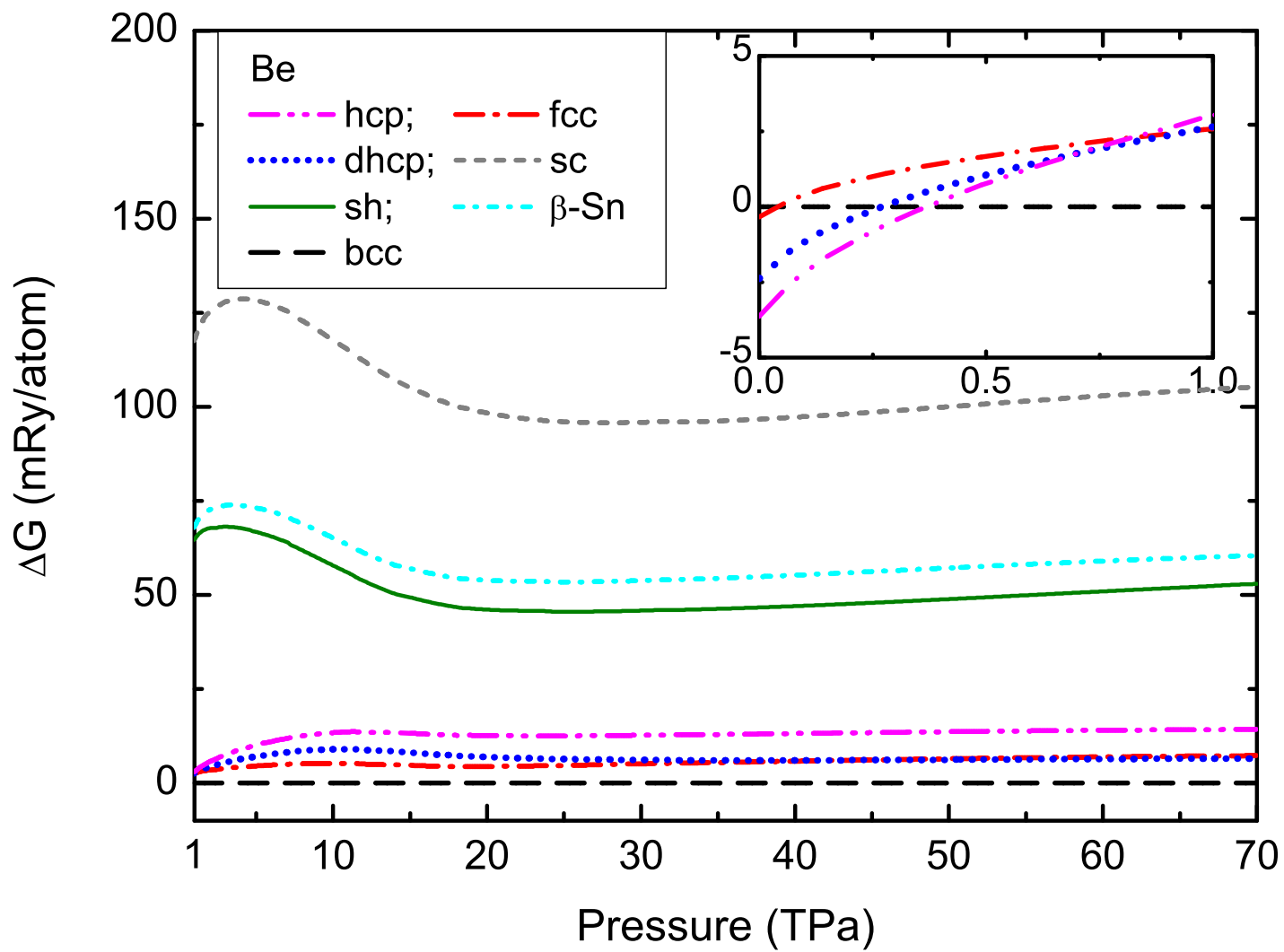}
  \caption{Gibbs energy difference for different beryllium phases at high pressures and zero temperature. The inset shows the pressure region below 1 TPa.}
  \label{fig2}
\end{figure}

\begin{figure}
  \includegraphics[width=\linewidth]{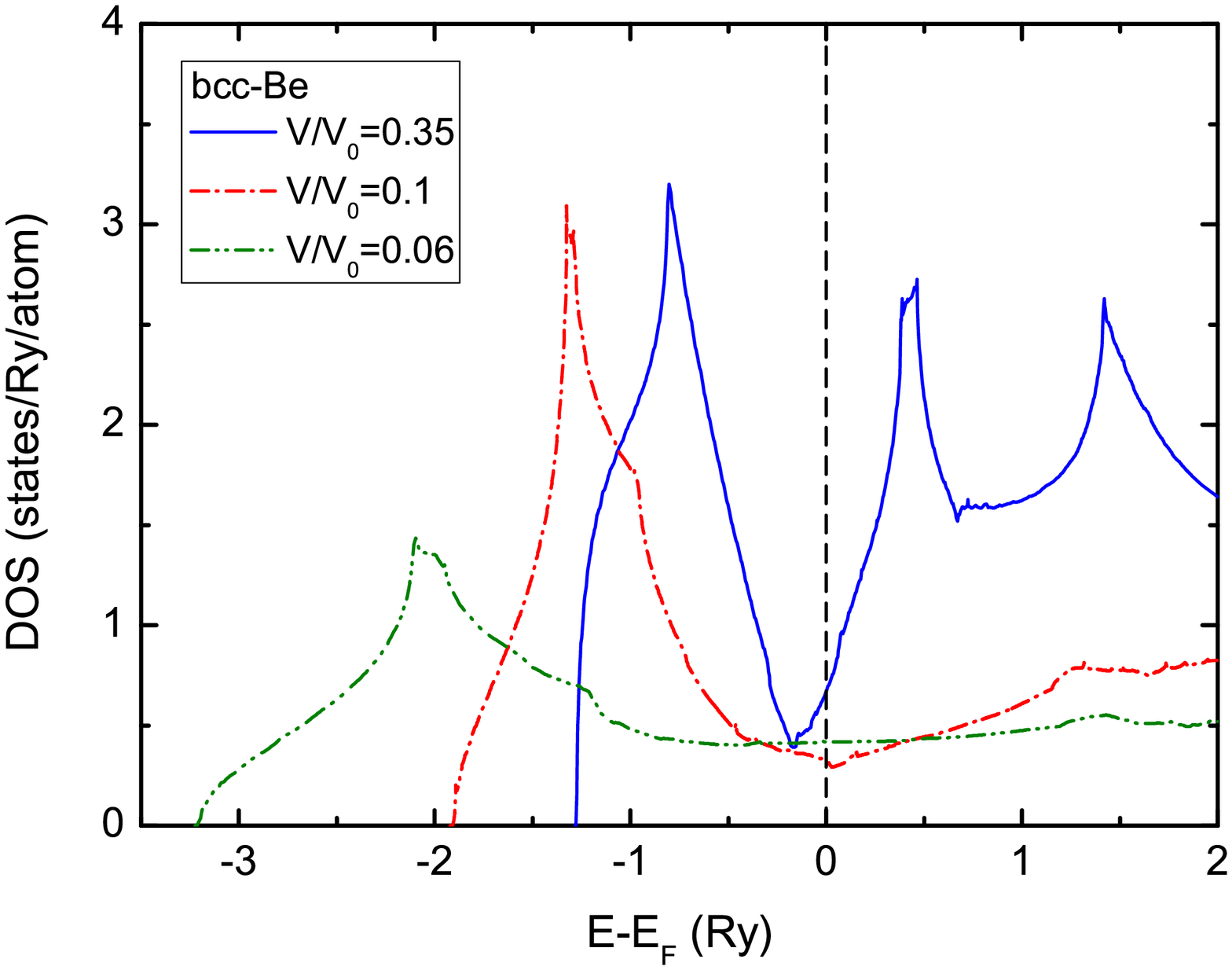}
  \caption{Electronic density of states of bcc beryllium for several compressions ($T$$=$0 K) corresponding to pressures $P$$\approx$0.66 TPa ($V/V_0$$=$0.35); $P$$\approx$12 TPa ($V/V_0$=0.1); and $P$$\approx$32 TPa ($V/V_0$$=$0.06).}
  \label{fig3}
\end{figure}

Consider first the relative stability of beryllium and magnesium structures of interest at $T$$=$0 K. Figure~\ref{fig2} presents results obtained in this work for beryllium. Hereafter Gibbs thermodynamic potentials versus pressure are presented relative to the bcc potential. As seen from the figure, hcp beryllium is thermodynamically most favorable at $P$ below 0.4 TPa which agrees well with other calculations \cite{A09,A14} and experiments \cite{A11,A41}. At higher pressures the bcc structure becomes energetically more favorable. It can be seen that further compression does not lead to any other structural changes. Our studies show the situation to remain unchanged at least up to $P$$\sim$250 TPa. Calculations also show the bcc phase to remain dynamically stable. It is well seen from Figure~\ref{fig2} that above about 20 TPa the slope of the curves $\Delta G$ weakly changes with the increasing pressure - all curves are almost parallel; the electron spectrum of Be stops changing significantly, demonstrating no fundamental changes under further compression. This can be seen in Figure~\ref{fig3} that shows the evolution of the electronic density of states of beryllium at high pressures. At $V$/$V_0$$=$0.35, a pseudogap (a large depression in the electron density of states near $E_F$) is seen to be still present in the electronic spectrum of Be. As mentioned in paper \cite{A05}, $s$-$p$ hybridization changes the electronic structure of Be so that the density of $s$-states on the Fermi level becomes virtually zero and the small number of the Fermi level crossing bands are mostly $p$-bands. It is however seen from Fig.~\ref{fig3} that the pseudogap gradually disappears under higher compression and the DOS becomes more and more similar to the nearly-free-electron density of states. This means that the above effect stops to manifest itself.

\begin{figure}
  \includegraphics[width=\linewidth]{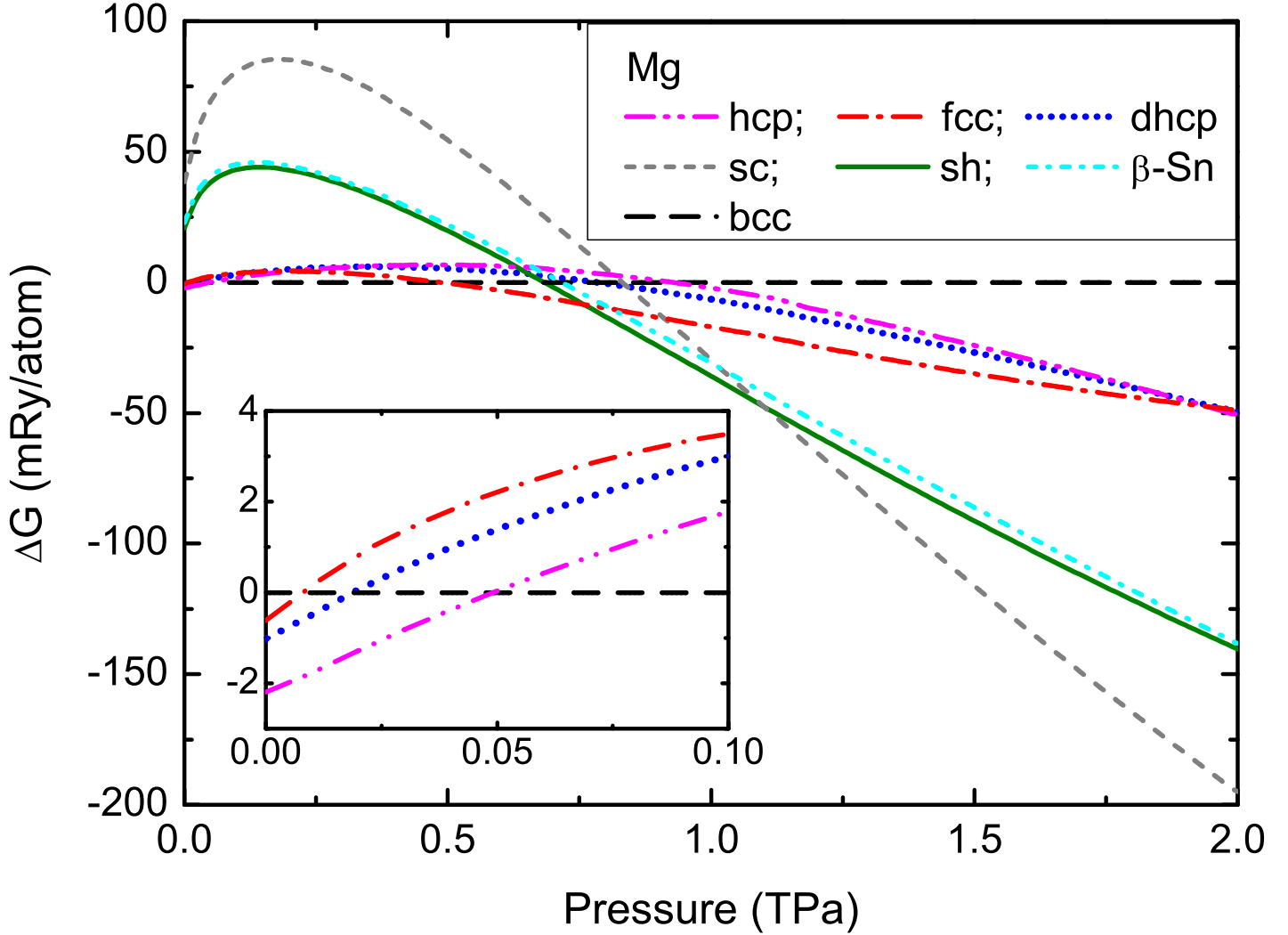}
  \caption{Gibbs energy difference for different magnesium phases under high pressures and zero temperature. The inset shows the region of low pressures below 0.1 TPa.}
  \label{fig4}
\end{figure}

As mentioned earlier, \emph{ab initio} results show magnesium is more diverse in structural changes under pressure than beryllium. Figure~\ref{fig4} shows the relative difference of thermodynamic potentials calculated for magnesium structures of interest at $T$$=$0 K. The hcp phase is thermodynamically most favorable up to pressures about 0.05 TPa which agrees with available experimental data \cite{A20,A42}. Then the structural transformations bcc-fcc-sh-sc are seen to occur at pressures about 0.48, 0.76, and 1.1 TPa, respectively. These values agree quite well with other \emph{ab initio} calculations \cite{A22,A27}. Our calculations suggest that the simple cubic structure of Mg remains thermodynamically most favourable and dynamically stable at least to a pressure of 12 TPa. It is seen from Figure~\ref{fig4} that in energy the $\beta$-Sn structure is quite close to the simple hexagonal phase but it does not become energetically more favorable anywhere in the pressure range under study.

\begin{figure}
  \includegraphics[width=\linewidth]{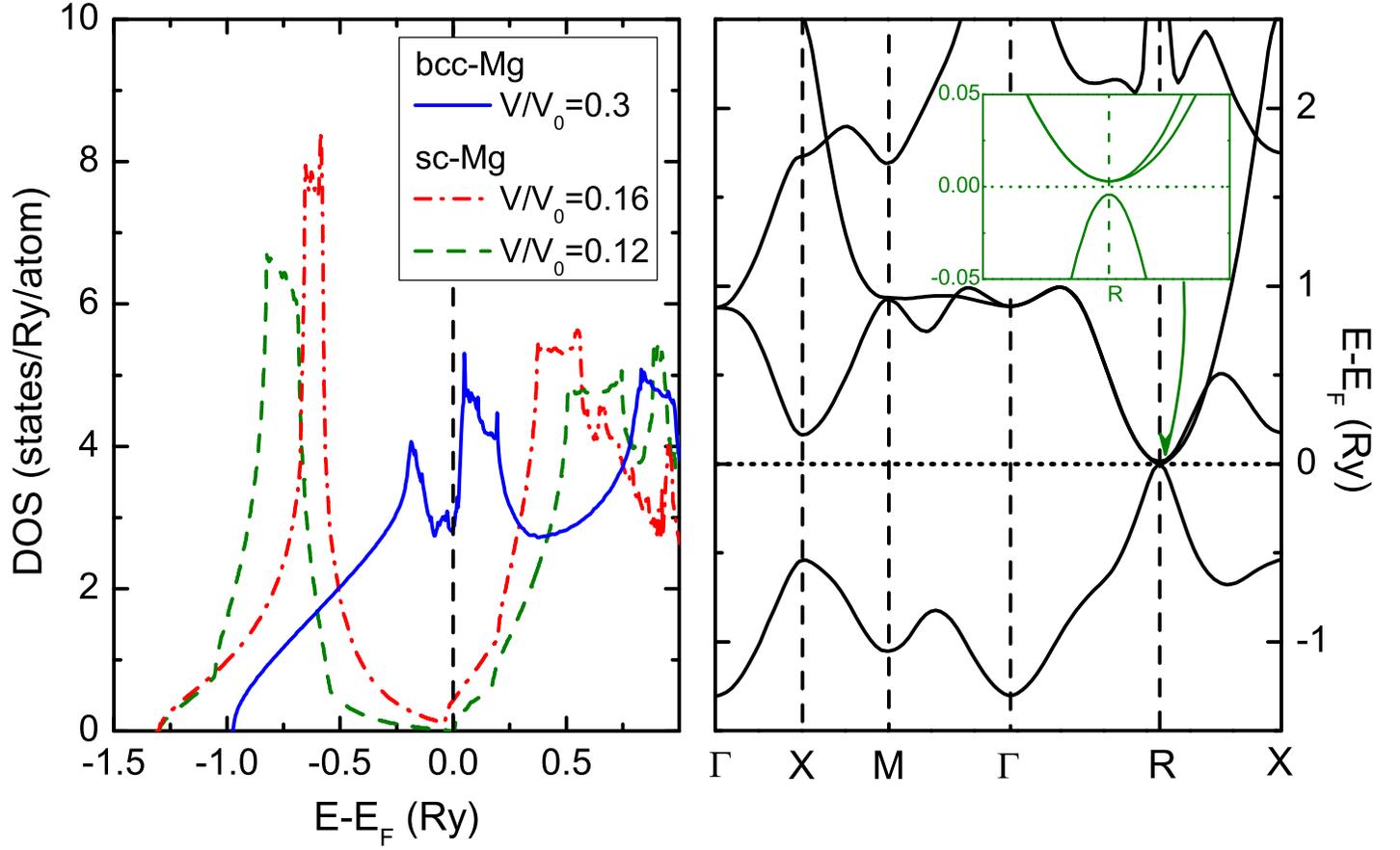}
  \caption{Electronic densities of states of bcc and sc magnesium at several compressions ($T$$=$0 K) corresponding to the following pressures: $P$$\approx$0.43 TPa ($V/V_0$$=$0.3); $P$$\approx$1.72 TPa ($V/V_0$$=$0.16); and $P$$\approx$3.36 TPa ($V/V_0$$=$0.12).}
  \label{fig5}
\end{figure}

Structural transformations in magnesium are accompanied by significant changes in the electronic spectrum. Figure~\ref{fig5} presents the evolution of the electronic DOS at zero temperature for different compressions, and a band structure of Mg at $V$/$V_0$$=$0.12. It is seen from the figure that at $V/V_0$$=$0.3 the electron density of states looks very much like the DOS of nearly-free electrons but further compression results in the appearance of a pseudogap that increases as $P$ grows. Finally, at a pressure of about 2.7 TPa, a narrow band gap ($<$0.1 eV) appears in the spectrum of valence electrons of the most stable cubic structure and magnesium becomes a semiconductor (see $V/V_0$$=$0.12). This is seen clearly in the Fig.~\ref{fig5} (the right panel) demonstrating the band structure of magnesium. It should be noted here that in theoretical paper \cite{A22}, one can notice a pseudogap in the electronic DOS for compressed magnesium but the authors did not consider pressures above 1 TPa and failed to detect the formation of the small energy gap. Note that the conventional density functional methods are known to underestimate the width of the band gap; it can be determined more accurately with the Green's-function theory (GW calculations) \cite{A43}.

\emph{Ab initio} calculations \cite{A44,A45} show a narrow band gap to also form in the electronic spectra of calcium and strontium under compression at zero temperature but much lower pressures. In fcc and sc Ca, it appears at $P$$\approx$30 GPa. However this state only exists at pressures between about 30 and 40 GPa. In the electronic spectrum of fcc strontium, the gap forms at $P$ below 3.5 GPa and disappears as the pressure slightly increases. Experiments show that the electrical resistivity of calcium and strontium significantly grows in the corresponding pressure ranges \cite{A44}. Under these pressures they both are semimetals with a very low concentration of charge carriers. Our calculations show that for magnesium, the pressure of transition into a semiconducting state is much higher and the interval of its existence is significantly larger, from $\sim$2.7 to 8.6 TPa, than in other alkaline-earth metals. Worthy of noting is one more similarity to calcium. A band gap also appears in fcc Mg but this structure is not thermodynamically most favorable in the above pressure range.

\begin{figure}
  \includegraphics[width=\linewidth]{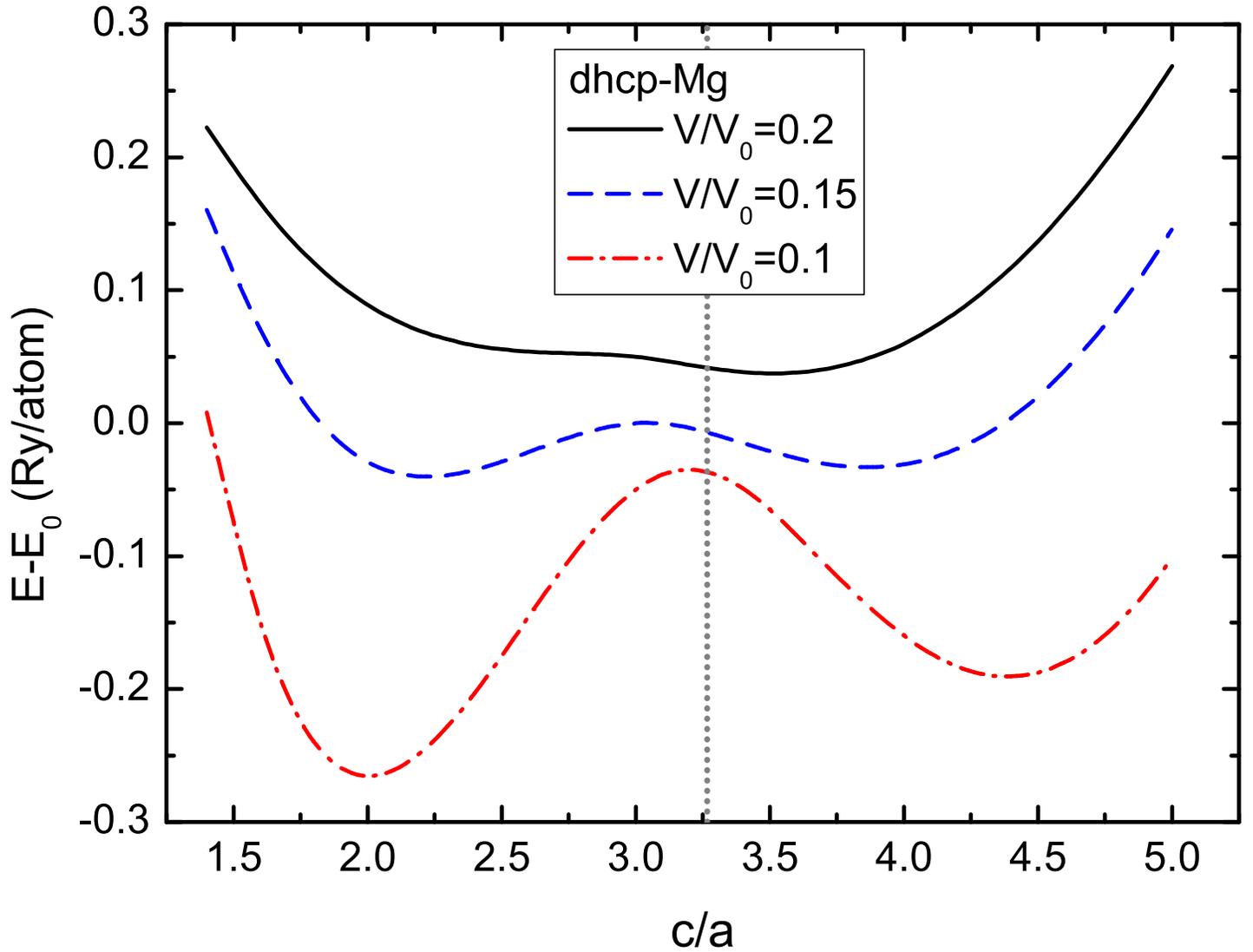}
  \caption{Internal energy of dhcp Mg versus $c/a$ at $T$$=$0 K for relative specific volumes $V/V_0$$=$0.2 ($P$$\approx$1 TPa), $V/V_0$$=$0.15 ($P$$\approx$2 TPa), $V/V_0$$=$0.1 ($P$$\approx$5.1 TPa). The vertical dotted line marks the ideal $c/a$$=$3.266.}
  \label{fig6}
\end{figure}

Experimental data \cite{A43} show that the magnesium's neighbor in the periodic table, sodium, becomes an optically transparent insulator at $P$$\sim$0.2 TPa. For a band gap to form in sodium, 5-fold compression is required \cite{A43}, while the semiconducting state of magnesium is reached at about 7.5-fold compression. Sodium has a dhcp structure at $P$$\geq$0.2 TPa with an unexpectedly low $c/a$ ratio $\approx$1.4 which is much smaller than the ideal value, 3.266. The dhcp structure of magnesium is not thermodynamically most favorable, but if we note how its energy depends on $c/a$ at different compressions, we can easily see analogies with sodium. Figure~\ref{fig6} presents internal energy versus $c/a$ for several compressions of dhcp magnesium. It is seen that at $V/V_0$$=$0.2 the curve $E(c/a)$ has only one energy minimum rather close to the ideal value of $c/a$ (the vertical dotted line). However, another minimum corresponding to low $c/a$$<$3 appears on the curve as compression increases. At $V/V_0$$=$0.1 the minima correspond to $c/a$$\approx$2 and $\approx$4.4. On whole, the behavior of $E(c/a)$ under growing compression is very much similar to what is observed for sodium (see Supplemental Material of Ref. \cite{A43}). But for magnesium, the appearance of the band gap is not observed for either of the two minima of the dhcp structure.

It can thus be stated that only beryllium under high and ultrahigh pressures behaves very much differently from other alkaline-earth metals. Exhibiting some unique features under ambient conditions, it becomes more and more similar to an ordinary simple metal as compression increases, whereas the behavior of magnesium under pressure looks like that of the heavy alkaline-earth metals. At pressures above 0.7 TPa, transformations into open structures occur in Mg, the crystal packing factor markedly decreases ($\sim$50\% for the sh phase), the electride structure appears \cite{A22,A27}, and the spectrum of valence electrons eventually change so that a narrow band gap forms in it at $P$$>$2.5 TPa.

\begin{figure}
  \includegraphics[width=\linewidth]{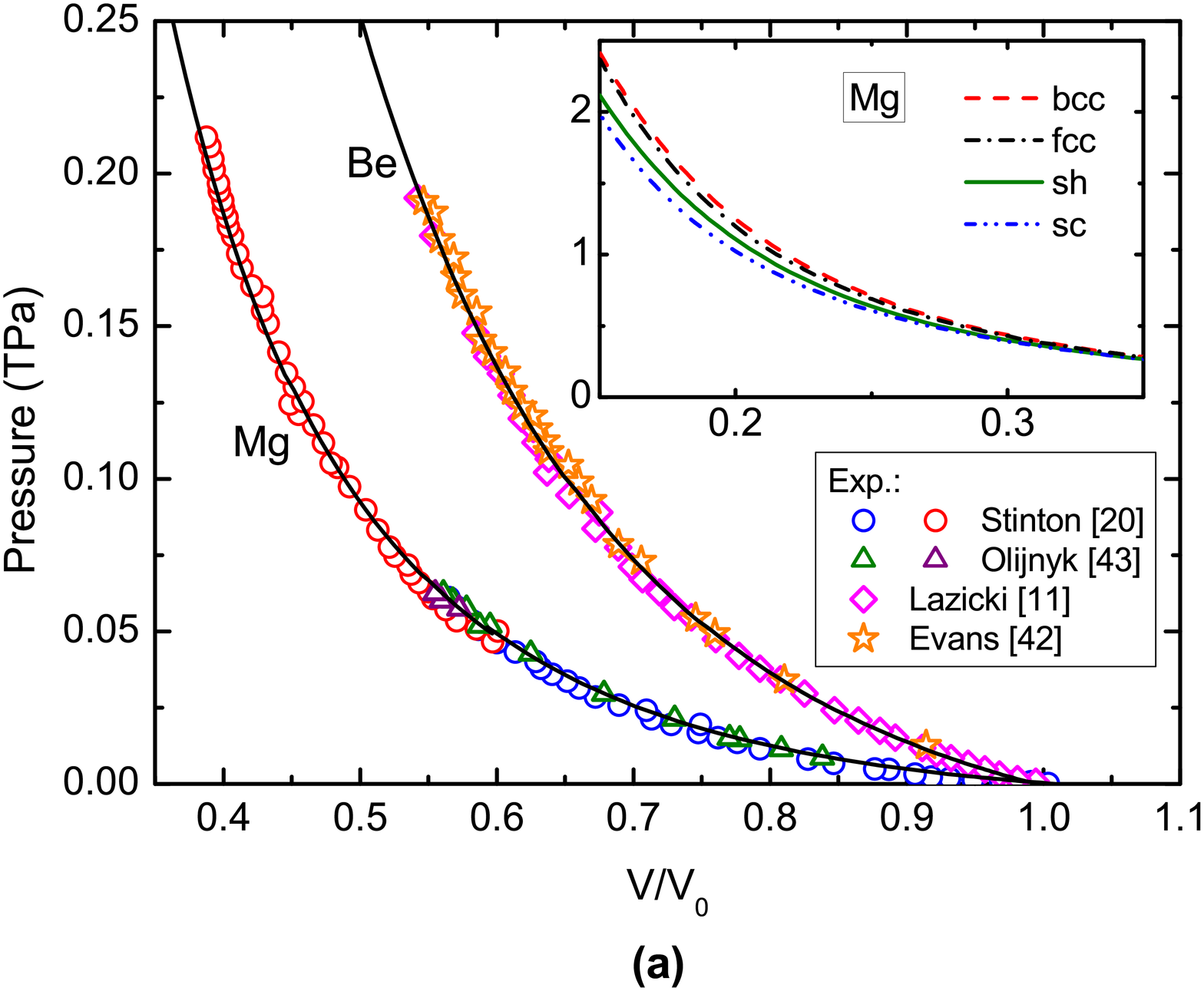}
  \caption{Pressure versus relative specific volume at room temperature for Be and Mg in comparison with experiment. Calculated results are shown by lines and experimental data are shown by symbols: stars \cite{A41}, diamonds \cite{A11}; triangles \cite{A42}, and circles \cite{A20}. The isotherm calculated for Mg is given with account for the hcp-bcc transition. The inset shows the ultrahigh pressure region with calculations for several Mg phases.}
  \label{fig7}
\end{figure}

Consider further the compression isotherms of beryllium and magnesium. Figure~\ref{fig7} shows calculated dependencies of pressure versus specific volume at room temperature in comparison with available experimental data. The curves calculated for both metals are seen to agree well with experiment. The inset shows the cold pressure of some magnesium structures in the higher compression region. As seen from the inset of Fig.~\ref{fig7}, the curves almost coincide at $V/V_0$$>$0.3, but with growing compression the difference in $P$ between the fcc, bcc structures and the open phases sh, sc of Mg markedly increases. This behavior is caused by electronic band rearrangement which makes the energy and pressure of the sh and sc structures lower compared to the other phases. Compressibility of open structures noticeably increases. A similar situation is observed in light alkali metals \cite{A46}, whose open structures also become energetically most favorable as compression increases.

\begin{figure}
  \includegraphics[width=\linewidth]{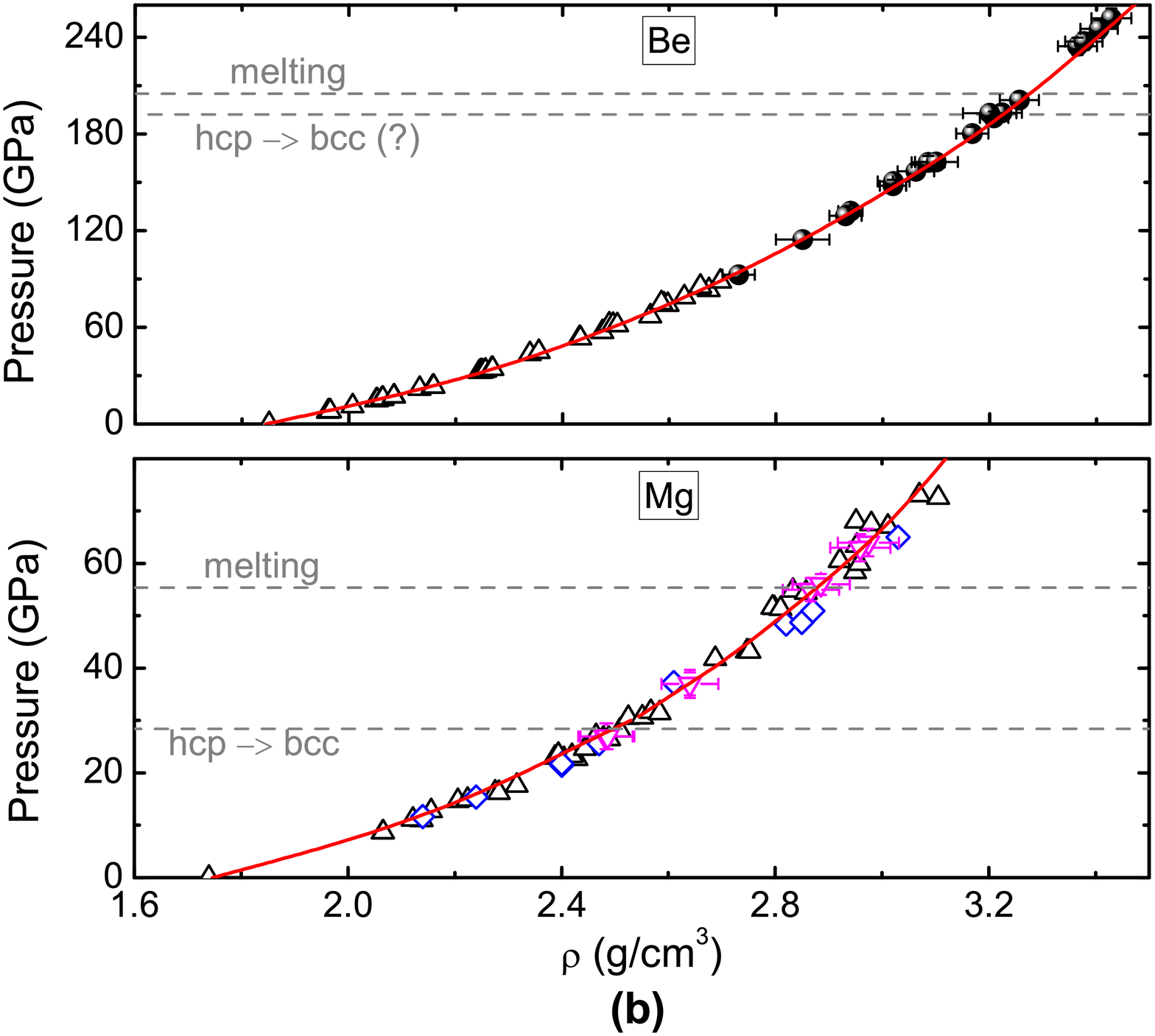}
  \caption{Shock Hugoniots of beryllium and magnesium in $P$-$\rho$ coordinates from our calculations (red lines) and experiments: triangles \cite{A47}, diamonds \cite{A48}, inverse triangles \cite{A49}, and circles \cite{A16}. The dashed lines show approximate boundaries of the hcp$\rightarrow$bcc transition and melting under shock compression.}
  \label{fig8}
\end{figure}

Now consider shock wave compression of beryllium and magnesium. Additional experimental data \cite{A16,A49,A50,A51} which help better understand their behavior under these conditions have recently appeared. Figure~\ref{fig8} compares the shock Hugoniots of Be and Mg from our calculations with data from different experiments. The horizontal lines mark the approximate pressures of the hcp$\rightarrow$bcc transition and melting. It should be noted that the presence of this transition in shock compressed beryllium remains debatable. It is seen that the calculations reproduce the shock compression of the two metals very well. The change of volume due to the transition is small and not seen in the pictures.

\begin{figure}
  \includegraphics[width=\linewidth]{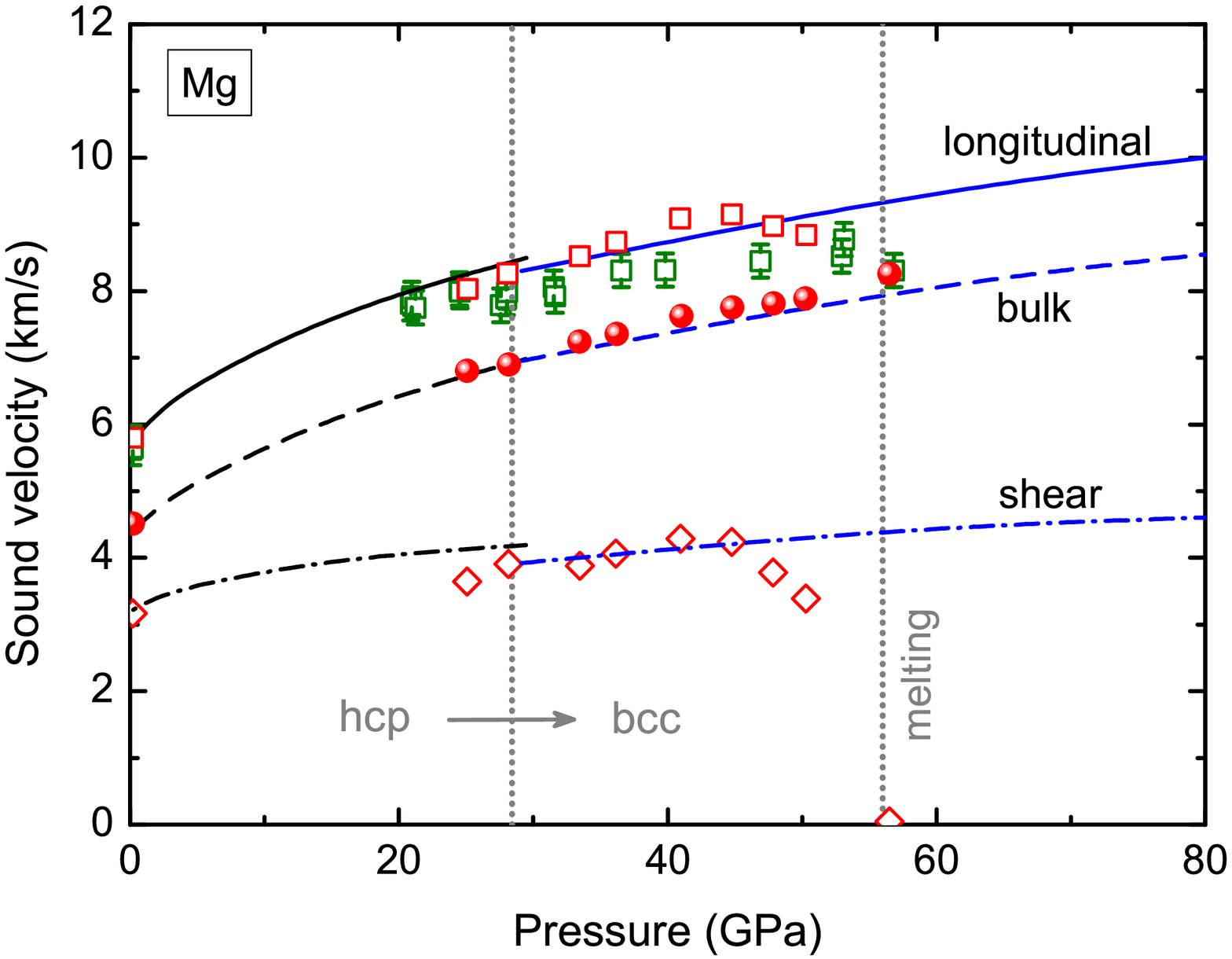}
  \caption{Magnesium sound velocity versus pressure under shock loading from calculation of elastic constants \cite{A21}, and from experiments: green squares \cite{A51}, red symbols \cite{A50}. The vertical lines mark the approximate pressures of the hcp-bcc transition and melting on the Hugoniot \cite{A49}.}
  \label{fig9}
\end{figure}

Earlier in paper \cite{A21} we calculated the elastic constants of several magnesium structures for different compressions at $T$$=$0 K. With these monocrystal constants we calculated longitudinal, shear and bulk sound velocities as functions of applied pressure for magnesium polycrystals using Hill averaging \cite{A52}. Figure~\ref{fig9} compares our calculations with recent measurements of sound velocities of Mg on the Hugoniot \cite{A50,A51}. The calculated and experimental data are seen to agree well. The hcp$\rightarrow$bcc transition results in a small jump of sound velocities which can hardly be detected in experiment with the available measurement accuracy. Since our calculations were done for zero temperature, they do not reproduce the softening of longitudinal and shear sound velocities while approaching melting.

\begin{figure}
  \includegraphics[width=\linewidth]{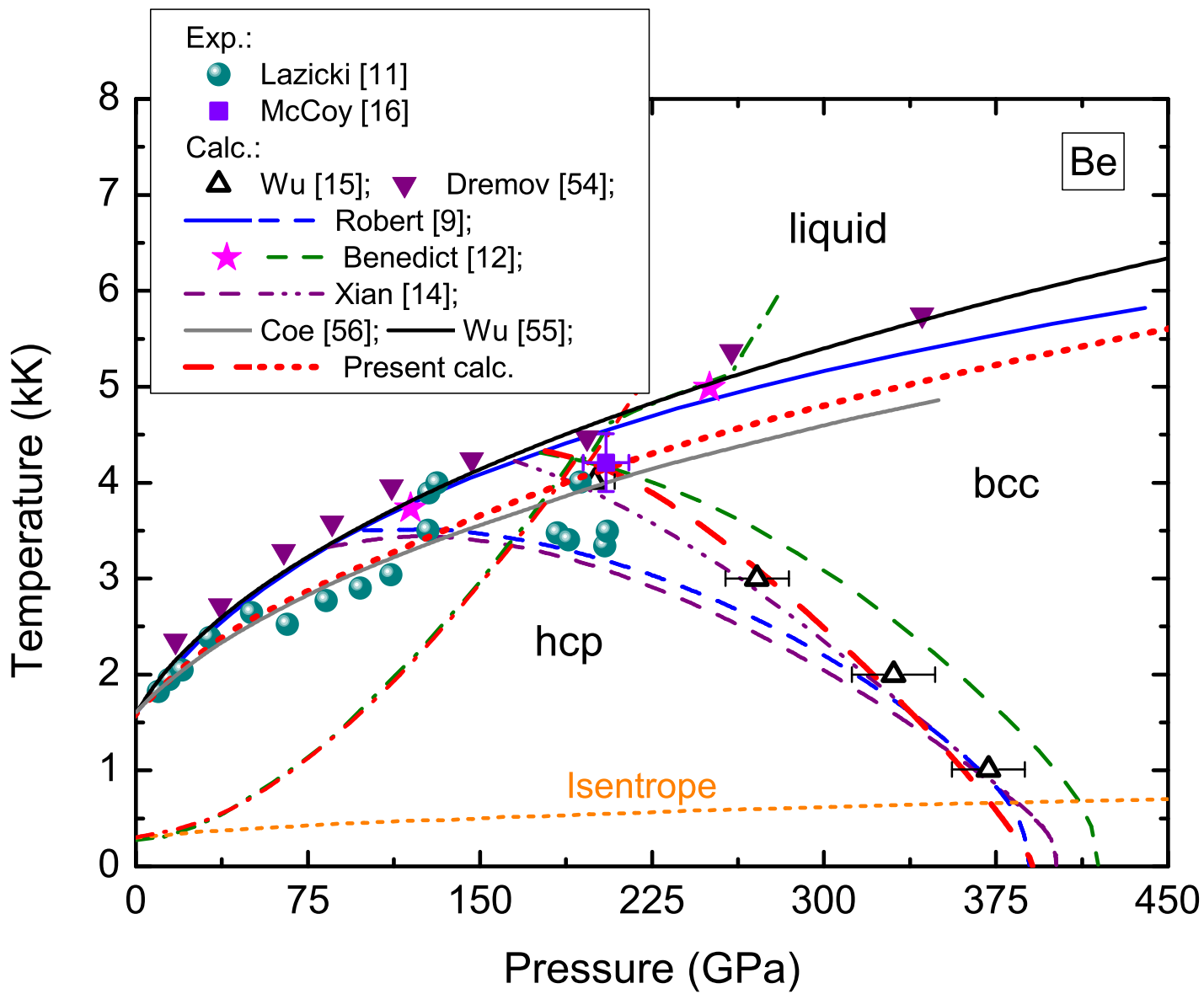}
  \caption{$PT$-diagram of beryllium. Red lines show calculations of this work (QHA): the dashed one is the hcp-bcc boundary; the dash-dotted one is the Hugoniot, and the dotted line is the melting curve. The hcp-bcc boundaries from other calculations are shown by the dashed blue line for calculated data from \cite{A09} (QHA); the dashed green line for data from \cite{A12} (QHA); the dashed and dash-dot-dotted magenta lines for calculations from \cite{A14} in QHA approximation and with full anharmonicity, respectively; open triangles for calculations with full anharmonicity from \cite{A15}. Melting curves from QMD calculations are shown by the blue line \cite{A09}, stars \cite{A12}, closed inverse triangles \cite{A53}, and the black line \cite{A54}. The gray line shows the melting curve obtained with a multiphase EOS \cite{A55}. The circles show data from static experiment \cite{A11}, marking the boundary of the region where the hcp phase exists. The square shows the estimation of the experimental melting point on the Hugoniot \cite{A16}. The principal isentrope was taken from the SESAME 2010 EOS \cite{A56}.}
  \label{fig10}
\end{figure}

Figure~\ref{fig10} presents a $PT$-diagram of beryllium calculated in this work in comparison with other \emph{ab initio} calculations and experiment. The issue of whether a pocket of the bcc phase exists at low pressures and high temperatures \cite{A09,A10} remained beyond the scope of this work. Circles in Fig.~\ref{fig10} mark the possible boundary of the region where the hcp phase of beryllium exists according to static experiment data \cite{A11}. Look first at the melting curve. For more correct comparison, the melting curve obtained for beryllium from the Lindemann criterion was calculated with the Debye temperature determined by the logarithmic phonon moment, as it is done in paper \cite{A54}. As seen from Fig.~\ref{fig10}, our curve underestimates the melting point with the increasing pressure compared to QMD calculations. Also, the hcp existence region from experiment \cite{A11} extends in temperature a bit higher than our curve runs. However it unexpectedly well agrees with the estimated melting point of Be on the Hugoniot from experiment \cite{A16}. It must be noted here that the authors of paper \cite{A53} suggested that prior to melting on the Hugoniot beryllium possibly transforms into the amorphous state which could explain some differences seen in Fig.~\ref{fig10} between the melting curves from MD calculations and experiment \cite{A16}. Despite that the Lindemann criterion and phonon spectra at zero temperature often give quite good melting curves \cite{A35,A39} that agree with experiment and MD calculations, here we are having some discrepancies. As shown in paper \cite{A54}, the melting temperature of beryllium from the Lindemann criterion at $P$$=$300 GPa is underestimated by 24\% compared to QMD calculations. Our calculations for this pressure give a smaller difference, 12\% only. Multiphase EOS calculations \cite{A55} where the melting curve is determined in the same way as we do, give a line close to ours (the gray line in Fig.~\ref{fig10}). Nevertheless, in further discussion of the $PT$-diagram of beryllium, we will rely on the melting curves of works \cite{A12,A54}, which are obtained more correctly from the physical point of view and agree very well with each other.

Consider the hcp-bcc boundary of Be. Fig.~\ref{fig10} presents these boundaries from calculations of two types. Results of the first type are obtained with the phonon spectra calculated at $T$$=$0 K and quasiharmonic approximation. The second type is QMD calculations which include all anharmonic effects. So, in paper \cite{A15}, the thermodynamic integration (TDI) method was used, while in Ref. \cite{A14}, phonon spectra were determined with account for temperature, using velocity autocorrelation functions. Most interesting here are results from \cite{A14} where the hcp-bcc phase boundary of Be was obtained in a unified manner for both types of calculations, QHA and QMD (magenta lines in Fig.~\ref{fig10}). They show that the contribution of anharmonic effects grows quite fast with the increasing temperature and at $T$$>$2 kK becomes essential for the determination of hcp-bcc boundary. An 'anharmonic curve' calculated in paper \cite{A14} agrees quite well with result from TDI calculations \cite{A15} (open triangles in Fig.~\ref{fig10}).

The situation with quasiharmonic calculations looks more intricate. Their results fall into two groups: with smaller \cite{A09,A14} and larger \cite{A12} (our work too) stability region of the hcp structure at high temperatures (magenta and blue dashes versus green and red ones). Our hcp-bcc boundary of Be (red dashed line) is steeper at high temperatures and at $T$$>$3 kK agrees well with results by Benedict et al \cite{A12}. Note that an analysis performed in Ref. \cite{A12} for the influence of anharmonicity effects on hcp and bcc phase stability, did not find them to contribute significantly which obviously disagrees with results from paper \cite{A14}. Data from our work and from \cite{A12} do not also contradict to static experiment \cite{A11} despite the use of quasiharmonic approximation in distinction from similar calculations presented in Refs. \cite{A09,A14}. It is quite challenging to understand what caused these discrepancies. The calculations were done with different codes but with one and the same exchange correlation functional PBE. Therefore the choice of functional is not a cause. There is a difference in approaches to quasiharmonic phonon spectra calculation for Be, specifically, supercells were used in Refs. \cite{A09,A14}, while the linear response method was used in our work and in paper \cite{A12}.

Now turn to the possibility of observing the hcp$\rightarrow$bcc transition in beryllium under shock compression. It is seen from Fig.~\ref{fig10} that the shock Hugoniot curves from our calculations and from [12] agree excellently. These calculations predict that the Hugoniot should be expected to cross the hcp-bcc boundary at $P$$\approx$191 GPa, $T$$\approx$4250 K. There is however a probability that the anharmonic effects will change the transition pressure at high temperatures and shift the boundary to higher pressures, as it was demonstrated in Ref. \cite{A14}. In this case the Hugoniot may not cross the hcp-bcc line. Then it becomes clear why no signs of a structural transformation are observed in dynamic experiment \cite{A16}. This is also in accordance with data from static experiment \cite{A11} where the bcc structure of Be was not detected.

\begin{figure}
  \includegraphics[width=\linewidth]{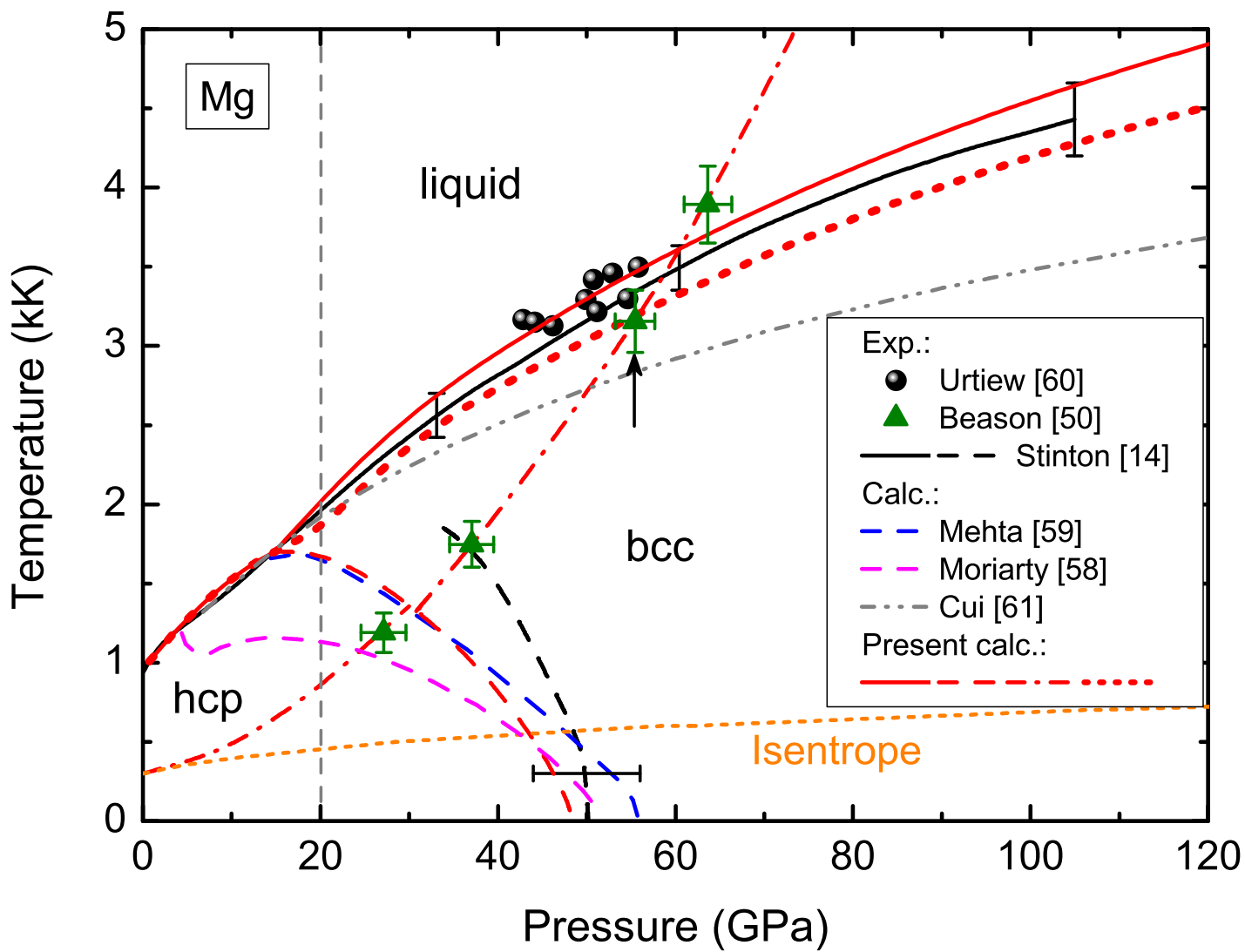}
  \caption{$PT$-diagram of magnesium. Red lines show results of this work (QHA): the dashed line shows the hcp-bcc boundary, the dash-dotted one is the Hugoniot, the solid line is the melting curve, and the dotted line is the melting curve for the hcp-bcc-liquid triple point at 20 GPa (the vertical gray dashed line, see the text). The orange dashed curve shows the principal isentrope from our calculations. Magenta and blue dashed lines show hcp-bcc boundaries calculated in Refs. \cite{A57} and \cite{A58} (GGA calculation), respectively. The black lines show data from static experiment \cite{A20}: the dashed line is the hcp-bcc boundary and the solid one is the melting curve. Triangles show points along the Hugoniot from experiment \cite{A49}, obtained with the Mie-Gruneisen EOS. The arrow points to the pressure corresponding to the onset of melting by data from \cite{A49}. The circles are points along the melting curve from dynamic experiments \cite{A59}. The gray dash-dot-dotted line is the melting curve from QMD calculation \cite{A60}.}
  \label{fig11}
\end{figure}

Let us now turn to the $PT$-diagram of magnesium. Figure~\ref{fig11} shows its phase diagram at relatively low pressures along with data from other calculations and experiments. Here all hcp$\rightarrow$bcc boundaries for Mg are from calculations in quasiharmonic approximation. Like in beryllium, the boundary has a negative slope. Our calculations agree well with calculations \cite{A58} at high temperatures, about 1 kK and above, and a bit worse at lower $T$. But they all are within the experimental error obtained in the determination of the transition pressure in experiment \cite{A19}. The curve calculated in work \cite{A57} underestimates the hcp phase stability region at high temperatures. At the same time static experiment \cite{A20} gives the steepest hcp-bcc boundary (the black dashed line in Fig.~\ref{fig11}) among all results. This may be indicative of the necessity to consider full anharmonicity in order to reproduce the slope of the phase boundary correctly. But shock-wave experiments \cite{A49,A50}, including those where the X-ray diffraction method is used for crystal structure analysis, give somewhat other results. So, the calculated Hugoniot shown in Fig.~\ref{fig11} is seen to cross the hcp-bcc boundary determined in static experiments \cite{A20} at $P$$\approx$37 GPa, while estimates from Ref. \cite{A50} give 28.4 GPa. Note here that our Hugoniot agrees very well with experimental points \cite{A49} determined with the Mie-Gruneisen EOS (green triangles in Fig.~\ref{fig11}). The value 28.4 GPa is close to the transition pressure from our calculations, 29.8 GPa. That is, theoretical studies here agree quite well with dynamic experiments but certain disagreement with static experiments is present.

The solid red line in Fig.~\ref{fig11} shows the melting curve $T_m$($P$) of Mg calculated in this work. It agrees quite well with data from static experiment \cite{A20} and shock experiment \cite{A59}. But the material of samples used in work \cite{A59} was not pure magnesium; that was an alloy with a magnesium content of 96\%. In more recent dynamic experiments \cite{A49,A50} where pure magnesium was studied, the onset of melting under shock conditions was detected at about 55.5 GPa. In Fig.~\ref{fig11} this pressure is pointed to by an arrow. The authors of paper \cite{A50} estimate the temperature at which shocked magnesium starts to melt to be about 3 kK which is somewhat lower than our estimate 3.4 kK. Since we use the Lindemann criterion, the trend of our melting curve is strongly dependent on the position of the hcp-bcc-liquid triple point from which it runs after the hcp-bcc transition. At this point calculations give a pressure of about 15 GPa. Estimation of experiment \cite{A50} gives $\sim$20 GPa (the vertical dashed line in Fig.~\ref{fig11}). By analogy with beryllium, it seems reasonable to suppose that the anharmonic effects may slightly shift the triple point to the high pressure region. In Fig.~\ref{fig11} we added a melting curve corresponding to a triple point at $P$$=$20 GPa (red dots). On the one hand, the new line does not contradict to static experiment \cite{A20} within its accuracy, and on the other hand, it moves us closer to the result obtained in paper \cite{A50}, taking into account also the finite error of these measurement. Our Hugoniot crosses this line at $P$$=$56 GPa and $T$$=$3.2 kK.

\begin{figure}
  \includegraphics[width=\linewidth]{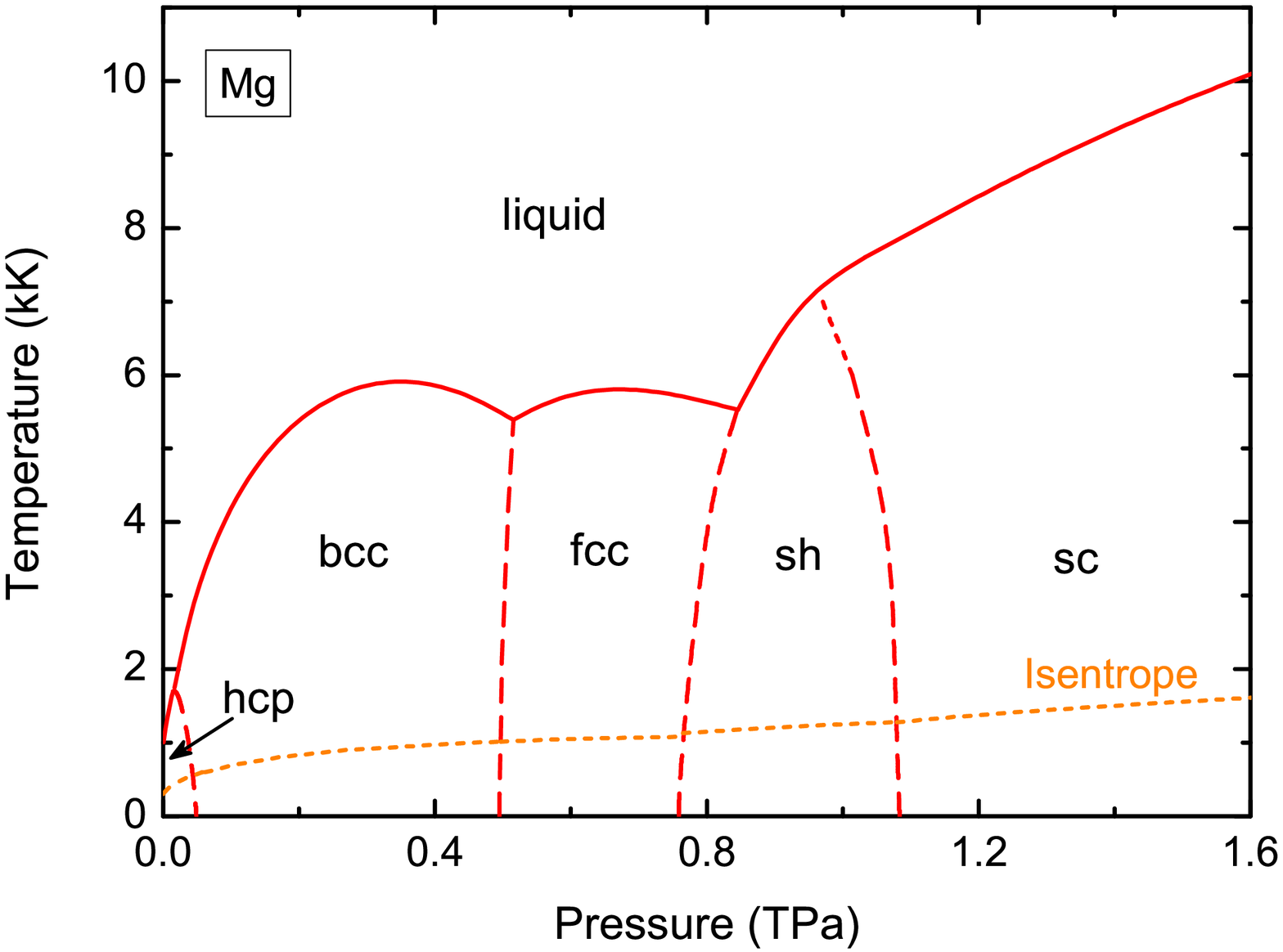}
  \caption{$PT$-diagram of magnesium calculated at ultrahigh pressures. The red solid line is the melting curve, the dashed ones are crystal phase boundaries, and the orange dashed line is the principal isentrope from our calculations.}
  \label{fig12}
\end{figure}

Figure~\ref{fig12} presents a phase diagram of magnesium obtained in this work for higher pressures up to 1.6 TPa. What draws attention here is the shape of the melting curve. It has small maxima at melting from bcc and fcc phases. The presence of a maximum on the bcc-liquid equilibrium curve for Mg is also confirmed by first-principles molecular dynamics calculations \cite{A60}. At pressures from $\sim$0.2 to $\sim$0.8 TPa, the melting curve changes weakly, but for the open structures sh and sc, $T_m$ is seen to steadily increase as pressure grows. This shape of the curve is similar to the $T_m$($P$) dependencies determined for calcium and strontium in experiment \cite{A61} where transition to open structures was accompanied by a noticeable growth of the melting curve after its more flattened previous part.

Characteristic maxima are known to be also present on the $T_m$($P$) curves of alkali metals (for example, see \cite{A62,A63}). As shown in our calculations, transition to a negative slope of the melting curve is caused by softening of some phonon modes in bcc and fcc magnesium as it happens, for example, in sodium and potassium \cite{A64}. So, for the bcc phase of Mg, the transverse modes $T_2$ and $T_1$ in the $\Gamma$N direction of the Brillouin zone sequentially undergo softening with increasing compression. This eventually leads to dynamic instability of the lattice and negative elastic constants $C'$$=$($C_{11}$-$C_{12}$)$/$2 and $C_{44}$. While for the fcc phase, only the transverse phonon mode in the $\Gamma$X direction softens and only the constant $C_{44}$ takes negative values.

Figure~\ref{fig12} also shows the principal isentrope of Mg calculated in this work. To calculate the principal isentrope, we determined the value of entropy $T$$S$(300 K)$=$$E$-$F$ under ambient conditions. The corresponding isentrope curve was determined from the condition $S$[$V$($P$,$T$),$T$]=$S$(300 K) for specified $V$ and $T$. The calculations were done in quasiharmonic approximation. Our calculations predict the isentrope to cross bcc-fcc, fcc-sh, and sh-sc boundaries at pressures $\sim$0.5, 0.76, and 1.08 TPa, respectively. It is worth noting that magnesium under pressure demonstrates quite a wide diversity of structural transformations and exhibits nonstandard physical properties for metal. State-of-the-art experimental ramp compression techniques \cite{A01,A02,A03} make it quite possible to detect the structural transition of interest even at very high pressures. So, very recent ramp experiments have confirmed the presence of simple hexagonal and cubic phases in magnesium at pressures above 0.8 TPa \cite{A65}.

\section{Conclusion}

In this work we have studied the structural properties of beryllium and magnesium under high and ultrahigh pressures using the FP-LMTO method for calculations. It is shown that magnesium under pressure demonstrate a wider diversity of structural transformations than beryllium. Besides the hcp$\rightarrow$bcc transition, three more transitions occur in Mg in the interval of pressures from 0.45 to 1.1 TPa. As a result of structural transformations in this metal, open phases, simple hexagonal and cubic, appear. The electronic spectrum of Mg changes strongly during compression, first a pseudogap forms in it, and then a narrow band gap appears at $P$$>$2.5 TPa, which indicates the onset of a semiconducting state. Magnesium under pressure behaves more similarly to heavier alkaline-earth metals, calcium and strontium. Beryllium under compression, on the contrary, gradually ceases to demonstrate unique properties. Its electronic spectrum becomes more and more similar to the spectrum of nearly-free electrons. After the hcp$\rightarrow$bcc transition at $P$$\sim$0.4 TPa and $T$$=$0 K, its structure remains unchanged up to very high pressures, at least to $\sim$250 TPa. The neighbors of beryllium and magnesium in the periodic table, lithium, sodium and aluminum, also demonstrate a large structural diversity under compression \cite{A27,A62,A63}. Thus, we can state here that beryllium as a crystal has one more peculiarity: it is very resistant to structural changes during compression in comparison with the other metals surrounding it in the periodic table.

% References
\medskip

\textbf{References}\\

\end{document}